# *Molecular dynamics simulation of silicon nanoparticle crystallization during laser-induced forward transfer printing*

Youwen Liang, Wan Shou[a)]

*Department of Mechanical Engineering, University of Arkansas, Fayetteville, AR 72701, USA*

a) Author to whom correspondence should be addressed. Email: wshou@uark.edu

**Abstract**

Laser-induced forward transfer (LIFT) printing is a versatile technique to realize micro/nano-scale additive manufacturing of functional materials, including metals and semiconductors. However, the crystallization phenomena during LIFT printing have not been well understood, which is critical to determine the resulting microstructure and properties. In this work, we systematically investigate silicon crystallization during LIFT printing using molecular dynamics (MD) simulations. Specifically, MD simulation with Stillinger-Weber (SW) potential is used to investigate the size effect and surface influence on the crystallization of Si nanoparticles during transportation in air. We found that with a decrease in nanoparticle size, crystallization becomes increasingly rare, even at low cooling rates. The nucleation location of different particles is also analyzed and almost always starts at a sub-surface location (below 5 Å). The evolution of the atomic structure during solidification is also monitored to guide LIFT printing of Si. Our simulation results indicate that nano-confinement induced by the surface layer can lead to single-crystal structure formation, which may shed light on additive manufacturing of single-crystal structures and devices.

## 1. Introduction

Silicon (Si), especially high-quality crystalline Si, has a wide range of applications in thin film transistors,[1,2] solar cells,[3] and other optoelectronic devices.[4] Although various manufacturing techniques have been developed for Si, additive manufacturing of Si is still underexplored.[1,5] In order to generate high-quality crystalline Si for devices, thermal-assisted crystallization is preferred, which typically requires a powerful energy source, such as pulsed laser irradiation,[6,7] thermal annealing [8], electron beam irradiation [9], and flash lamp annealing [10]. Among various power sources, lasers have shown great versatility in crystallization and patterning without a mask. Although additive manufacturing of Si followed by laser crystallization has been explored [5], direct printing of crystalline Si remains challenging and can be a promising approach for device fabrication. Recently, laser-induced forward transfer (LIFT) printing has emerged as a versatile technique for the precise deposition of various materials, including metals, polymers, and semiconductors.[11] Zywietz et al. and Makarov et al. showed that amorphous Si structures can be printed with LIFT.[12,13] However, the understanding of the Si transformation during printing is limited, which prevents us from controlling the crystallization process for high-performance crystalline Si printing.

Crystallization of amorphous Si has been extensively explored experimentally, but the fundamental mechanism of Si crystallization is not fully understood yet. In situ observation of the crystallization process is very limited due to the relatively high temperature and short time scale. Recently, Xiang et al. and In et al. reported in situ TEM observation of Si nano-pillar crystallization induced by laser melting and solidification on $SiO_2$.[14,15] This technique provided a powerful tool to gain real-time feedback on nanoscale crystal structure change. However, it remains challenging to visualize the atomic transformation and, correspondingly, measure localized physical quantities such as energy and temperature. In this situation, molecular dynamics (MD) simulation becomes an excellent approach to reveal the invisible atomic transport phenomena and quantify experimentally inaccessible parameters.[16–19] Lee et al. performed MD simulations of the excimer laser annealed Si nucleation and crystallization process on $SiO_2$ substrate,[16,17] and they found that the {111}-oriented nuclei at the surface were driven by the lower surface energy. Li et al.reported the surface-induced crystallization in Si and Ge;[19] they found that the free surface can accommodate the volume expansion and facilitate the nucleation. Among various factors, the size of silicon nanoparticle is a critical factor that affects the nucleation and crystallization.[20–22] Li et al. found that the Laplace pressure is one of the reasons that restrain water nanodroplets from crystallizing.[20] The system size has also been reported to affect the initial temperature and final atomic structure.[23–26] Another research by Lü et al. stated that the nucleation probability is related to the structural order wave, which is due to the variety of density from volume expansion. They found regions with high nucleation probability are located approximately 30–50% of the particle bulk region beneath the surface.[21] However, they did not provide a complete perspective of the whole crystallization process and only focused on the early stages (before a stable crystalline cluster forms). It is generally accepted that the surface can enhance nucleation.[19,27] Zhao et al. combined MD simulation of Si nanoparticle crystallization in an Ar atmosphere with experiment.[26] They concluded that the probability of forming a crystalline phase depends on the cluster formation temperature and the cooling rate. A latent heat release was observed in the case where crystallization occurred. Besides, single-crystal structures were obtained in their study. Yet, the details of crystal evolution were not studied.

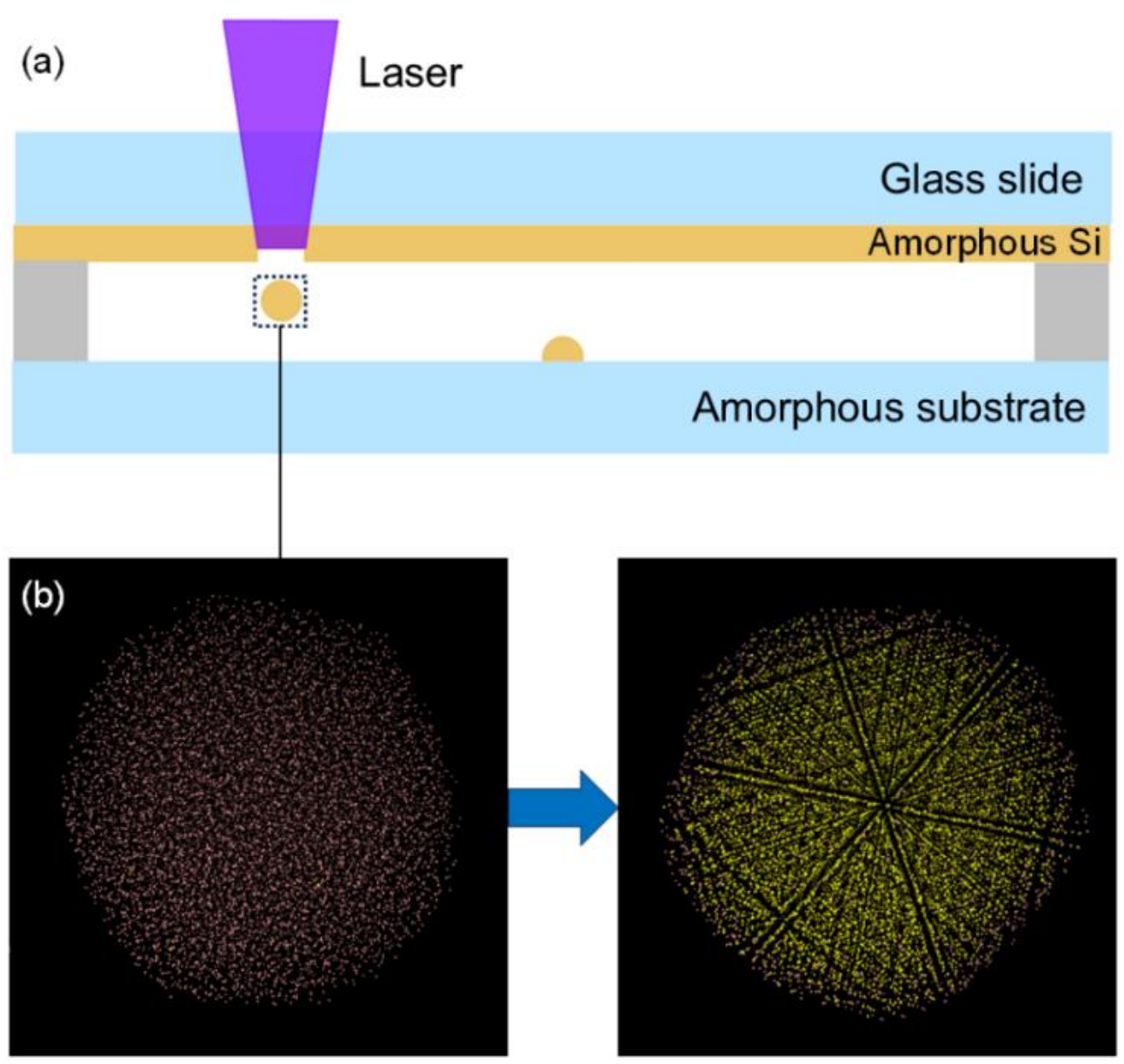

Fig. 1 (a) Illustration of the LIFT printing process. (b) Representative snapshots of molten Si nanoparticles crystallized into a single-crystal nanoparticle.

In this paper, we investigate the nucleation and crystallization evolution of Si nanodroplets during the transportation of LIFT in air with molecular dynamics (MD) simulations. Fig. 1 (a) is an illustration of the LIFT printing process for Si nanoparticles. The basic setting for a LIFT printing includes a donor (for example, a glass slide coated with amorphous Si), a substrate (for example, a clear glass slide), and a pulsed laser. The laser is focused on the amorphous Si, and then the coated Si will be melted and ejected from the donor. Ideally, the melted amorphous Si droplet can cool down from an amorphous liquid structure and crystallize into a single crystal structure during the transportation (namely, before the droplet hits the substrate). In this study, we focus on the melting and solidification of the Si nanodroplet after ejection from the donor but before hitting the substrate (as indicated in the dotted square in Fig. 1(a)). This process can simplify complex interactions with substrates, but provides a fundamentally important landscape of nucleation and crystal growth evolution. Fig. 1 (b) shows a promising scenario where the melted amorphous Si is crystallized to a single-crystal nanoparticle from our simulations.

## 2. Simulation method

LAMMPS was used to conduct MD simulations.[28] A well-established potential, the Stillinger-Weber (SW) potential, was used for the simulation of Si.[29] Silicon nanoparticles of different diameters (from 2 to 20 nm) were created in LAMMPS. All the simulations were conducted under a vacuum environment. The timestep of the simulation is 1 fs. The simulation box for nanoparticles between 2-12 nm is set to 16×16×16 $nm^3$, and for 14-20 nm is set to 27×27×27 $nm^3$. All Si nanoparticles with different diameters were heated up from 300 to 2000 K in 10 ns, and held at 2000 K for 5 ns to ensure the Si particle is completely melted into amorphous state. During the heating process, periodic boundary conditions, an isothermal canonical ensemble (NVT), and a Nose-Hoover thermostat were used. The melting points of different Si nanoparticles were determined based on the drastic changes in potential energy during heating.[30] For the solidification process, the molten Si nanoparticles obtained from the melting process were first stabilized at 1700 K and then, used as the input for cooling. During the solidification simulation, the micro-canonical ensemble (NVE) was used, so that the temperature evolution can represent the experimental condition.[26] The timestep and simulation box size were kept the same as in the melting process. The lowest temperature before the rise in temperature due to latent heat release, marked with a dashed circle in Fig. 3, was treated as the critical temperature of crystallization. The melting points and crystallization temperatures of different sizes are shown in Fig. 2. The number of atoms in Si nanoparticles is shown in Table 1.

Table 1. Number of atoms in different Si nanoparticles.

| Diameter (nm) | 2 | 4 | 6 | 8 | 10 |
|---|---|---|---|---|---|
| Number of atoms | 191 | 1707 | 5707 | 13407 | 26167 |

| Diameter (nm) | 12 | 14 | 16 | 18 | 20 |
|---|---|---|---|---|---|
| Number of atoms | 45215 | 71793 | 107161 | 152405 | 209121 |

The systems were cooled down to 800 K for the crystallization study. The same thermal conductance of $6.38 \times 10^5$ W/m$^2$K was applied to all the sizes to obtain the critical crystallization temperature as shown in Fig. 2 (b). The nanoparticles with diameters of 2, 4, 8, and 12 nm were focused to study the crystallization process with the same thermal conductance as shown in Fig. 3 (a). 20 cases were simulated for 2 nm and 4 nm nanoparticles, and 10 cases were simulated for 8 nm and 12 nm nanoparticles. Additionally, different thermal conductance was applied to the 8 nm Si nanoparticle to investigate the relationship between cooling rate and structure evolution.

The simulation results were visualized and analyzed by OVITO and VMD.[31,32] The atomic pressure and mean squared displacement (MSD) were calculated and output by LAMMPS. The radial distribution function (RDF) and coordination number were calculated by OVITO.

In order to identify crystalline clusters from liquid (or amorphous) atoms, the method of bond order parameters (BOP) was employed in this study.[25] BOP has been widely used for different materials, such as gold, carbon, and Si.[19,25,33] Specifically, here, local order parameter $q_3$,[19,34] which is sensitive to the crystalline order, was used for Si. The local structure around the atom $i$ is given by

$$\bar{q}_{lm}(i) = \frac{1}{N_b(i)}\sum_{j=1}^{N_b(i)} Y_{lm}(\theta(\vec{r}_{ij}), \emptyset(\vec{r}_{ij})),$$

where the sum runs over all $N_b(i)$ bonds of atom $i$, and θ and ∅ denote the azimuthal and polar angles of orientation for the bond $\vec{r}_{ij}$. By constructing a $2l+1$ dimensional vector $\vec{q}_l = [\bar{q}_{l,-l}, \bar{q}_{l,-l+1}, \ldots, \bar{q}_{l,l}]$, we can compute local invariants

$$q_l = \frac{1}{N_b(i)}\sum_{j=1}^{N_b(i)} \frac{\vec{q}_l(i)\cdot\vec{q}_l^*(j)}{|\vec{q}_l(i)|\,|\vec{q}_l^*(j)|}.$$

A cutoff distance of 0.293 nm was used to select the nearest neighbor,[29] and a cutoff value of $q_{3c} = -0.75$ was adopted to identify crystalline atoms from amorphous atoms.[19,34,35]

# 3. Results and discussion

## 3.1 Particle size and cooling rate influence

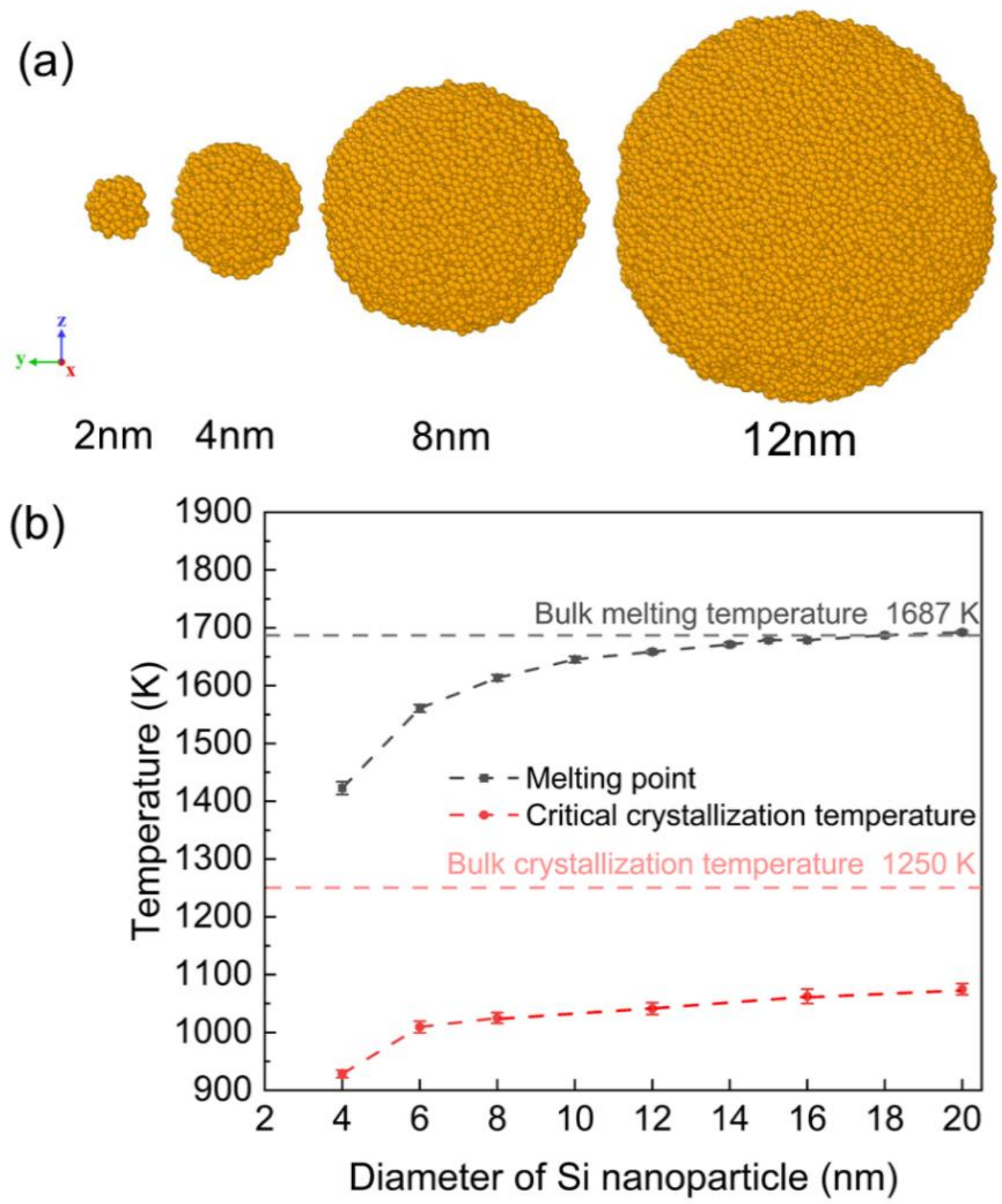

Fig. 2 (a) Snapshot of molten Si nanoparticles with different sizes; (b) Comparison of melting point and crystallization temperatures of Si nanoparticles with different sizes.

To investigate the particle size influence on crystallization as well as melting, Si nanoparticles with varying diameters are studied in the simulation. Four representative molten Si nanoparticles with diameters of 2, 4, 8, and 12 nm are shown in Fig. 2 (a), with a clear size difference. The melting point and crystallization temperatures of nanoparticles in the range of 4 nm to 20 nm are shown in Fig. 2 (b). The melting and crystallization temperatures of 2 nm are not included, as no clear potential energy change (used for determining melting point) or crystal formation was observed in our melting and solidification simulations (although with varying cooling rates, ranging from 0.1 to 5 K/ns between 1000 K and 800 K). The detailed potential energy evolution of 2 nm and 8 nm was shown in Fig. S1 in the supplemental information. It is obvious that the melting point increases with increasing diameter, while generally being lower than the bulk melting point. A similar trend is observed for the crystallization temperature, lower than the bulk crystallization temperature. It is also noticed that with the increasing size of nanoparticles, the melting point and crystallization temperature approach the corresponding bulk temperatures.[26,36]

Yet, due to the significant difference in surface between bulk and nanoparticles, we expect a significant difference in melting and crystallization.

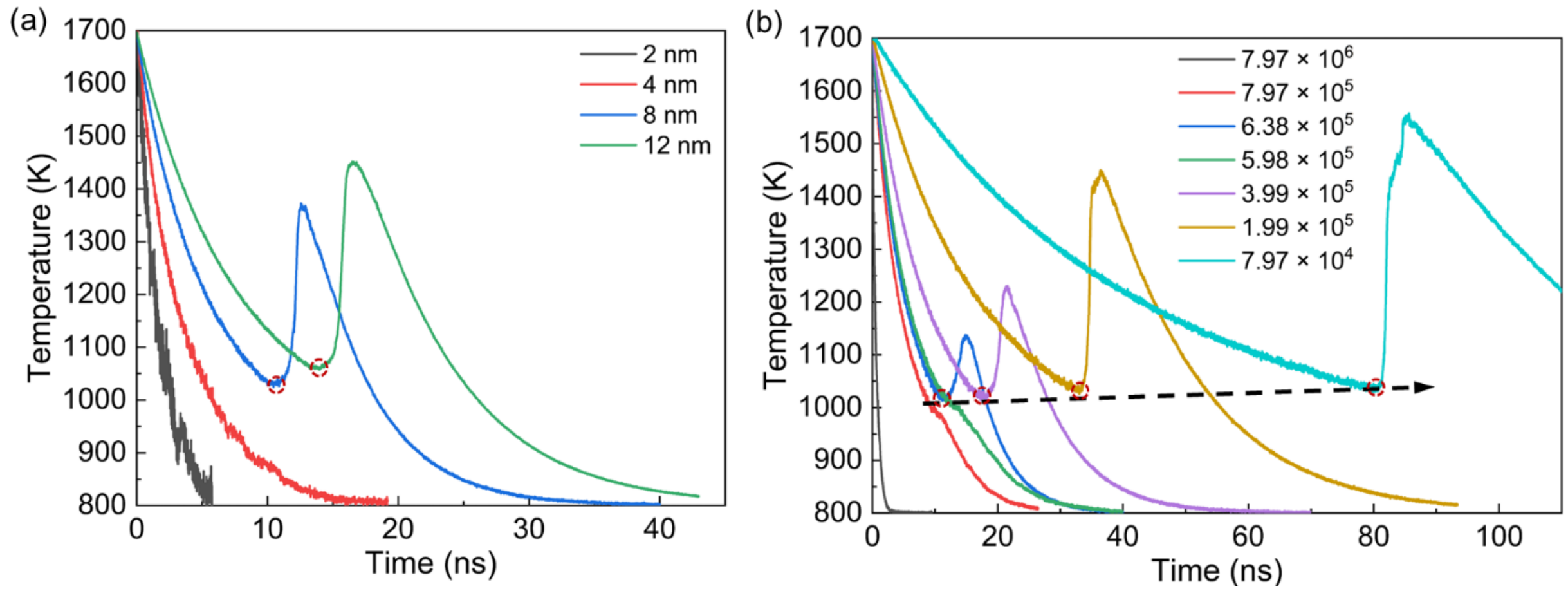

Fig. 3 Temperature evolution of Si nanoparticles: (a) Different diameters and (b) 8 nm particles with different thermal conductance ($W/m^2K$). (Note: in (a), the average cooling rates for 2 nm and 4 nm particles are about 108.15 K/ns and 43.3 K/ns from 1100 K to 900 K, and for 8 nm and 12 nm particles are 26.2 K/ns and 17.0 from 1100 K to the lowest temperature before latent heat release.)

To provide a better understanding of the size effect on temperature evolution during solidification, we plot the representative results of four different nanoparticles, cooled from 1700 K (as shown in Fig. 3a). Here, the same interfacial thermal conductance ($6.38 \times 10^5$ $W/m^2K$) is used for all four nanoparticles, resulting in a noticeable decrease in cooling rate as particle size increase. Due to the fast cooling rate (~108 K/ns and 43 K/ns), 2 and 4 nm Si nanoparticles did not show a clear temperature jump ($\Delta T$, i.e., latent heat release). In contrast, representative Si nanoparticles with 8 and 12 nm diameters showed a significant latent heat release during cooling, similar to previous reports.[26] It should be noted that with the decrease of the cooling rate, a 4 nm particle can successfully crystallize (see supplementary information Fig. S3). It is confirmed from our simulation that the release of latent heat is closely associated with crystallization (Fig. 5). Once crystallization begins, grain growth is fast, and the temperature jumps in the following several nanoseconds. However, the temperature jump is rarely observed for high thermal conductance cases (corresponding to a higher cooling rate). It is noted that earlier and steeper latent heat release typically is accompanied by an earlier 6-ring structure.[26] Furthermore, we show that the thermal conductance (corresponding to cooling rate) significantly influences the temperature evolution in Fig. 3 (b). As indicated by the dashed arrow, with the decrease in cooling rate, the crystallization temperature slightly increases; meanwhile, the total amount of released latent heat clearly increases from left to right (from fast to slow cooling). According to classical nucleation theory, $J = J_0 \exp(-\frac{\Delta G^*}{k_B T})$, within a certain range, lower temperature significantly enhances the nucleation rate,[34] which means a slow cooling rate (associated with a higher crystallization temperature) favors bigger grain formation (due to suppressed nucleation). As the increasing cooling rate leads to supercooling (both for smaller particles in Fig. 3 (a) and the left side of Fig. 3 (b)), crystallization may be suppressed, resulting in an amorphous structure. Therefore, control of the cooling rate is critical for crystalline Si printing.

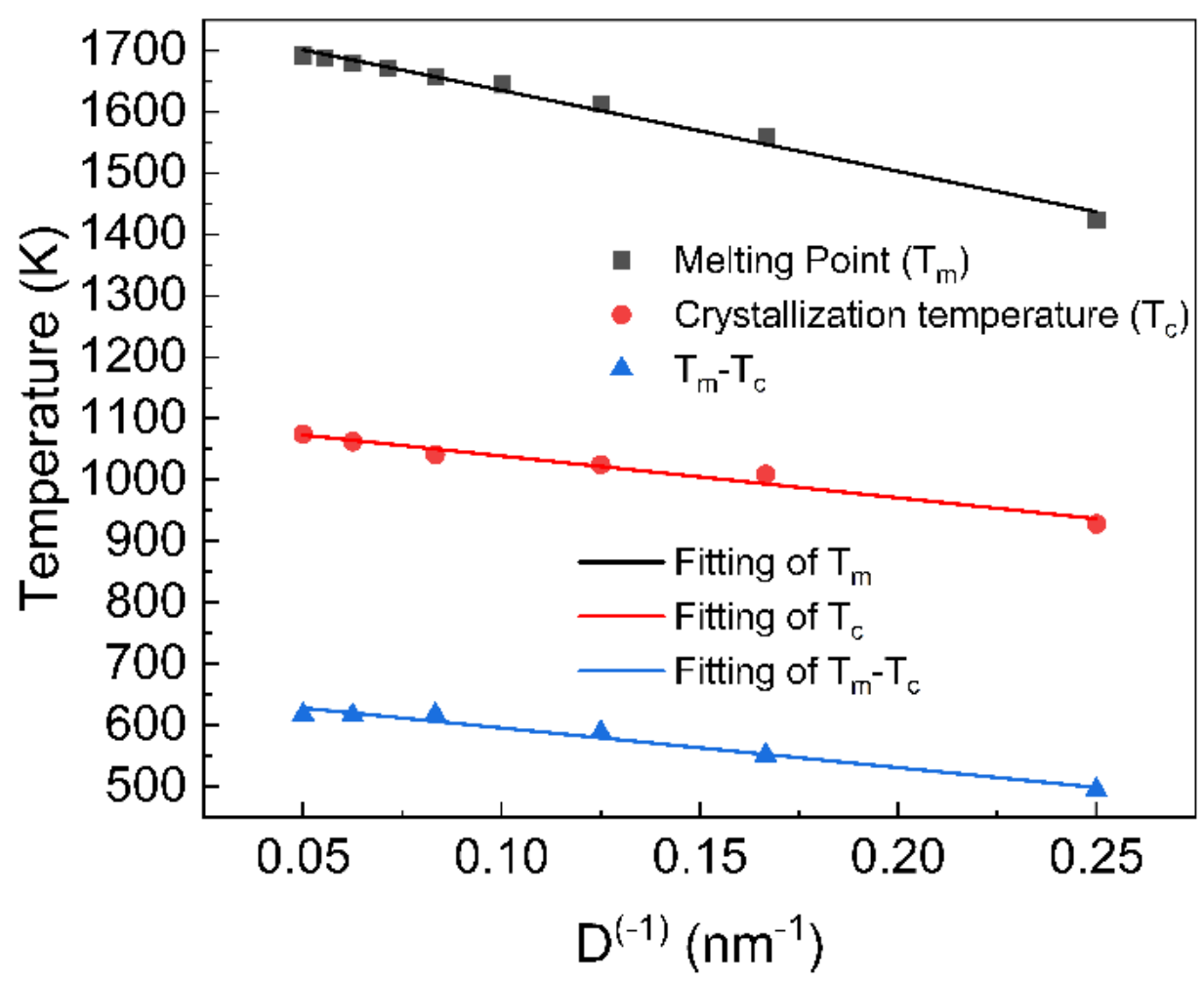

Fig. 4 The relationship between nanoparticle diameter and critical temperatures (melting point, crystallization temperature, and undercooling).

To further understand the relationship between the melting point ($T_m$), and crystallization temperature ($T_c$), and the particle size, we plot and fit the data points in Fig. 4. Similar to the previous report,[36] a linear relationship is observed between $T_m$ and the reciprocal of particle size (1/D)), namely, $T_m \propto D^{-1}$ (see supplementary information for the derivation). Interestingly, this linear relationship is also valid for crystallization temperature ($T_c$) and the largest undercooling ($\Delta T = T_m - T_c$), namely, $T_c \propto D^{-1}$, and $\Delta T \propto D^{-1}$. This means it is critical to control the droplet size (or curvature) during LIFT printing in order to tune the crystallization temperature and the subsequent solidification.

### 3.2 Atomic structure evolution during solidification

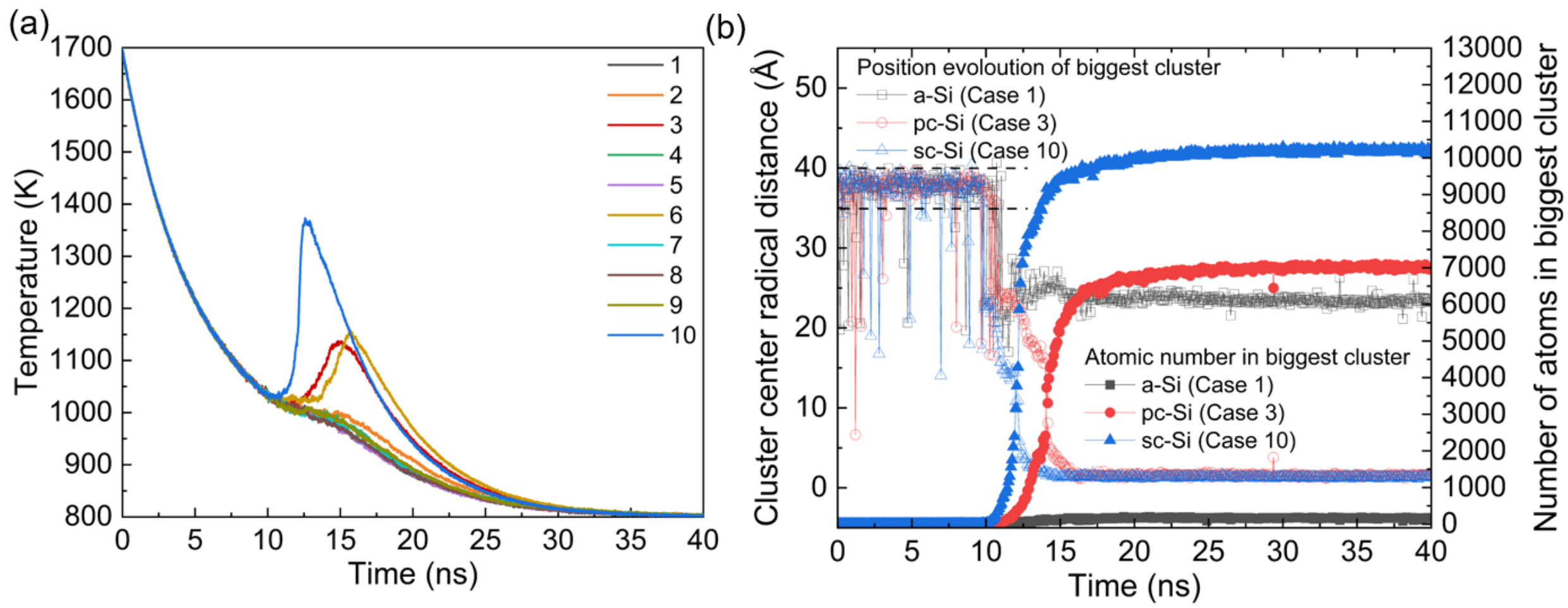

Fig. 5. Temperature and crystalline atom evolution of the 8 nm Si nanoparticle during crystallization. (a) Temperature evolution of 10 cases during cooling with the same thermal conductance ($6.38 \times 10^5$ W/m$^2$K). (b) The variation of crystalline atom number in the biggest cluster and the corresponding cluster center location in amorphous (case 1), polycrystalline (case 3), and single-crystalline (case 10).

To better probe the structural evolution, the 8 nm Si nanoparticle is further examined. We ran the simulations 10 times with the same thermal conductance with an average cooling rate (between 1200 K to 1100 K) of about 43 K/ns (but with different initial temperature distributions) and plotted their temperature history in Fig. 5 (a). Among these 10 cases, three cases show distinct temperature jumps. This means the crystallization at this cooling rate (or thermal conductance, $6.38 \times 10^5$ W/m$^2$K) is not certain. With a further decrease in thermal conductance to $1.99 \times 10^5$ W/m$^2$K (with an average cooling rate of about 13.25 K/ns), we ran 10 cases, and all of them exhibited stable crystallization (see supplementary information Fig. S2). It is believed that as long as a sufficiently low cooling rate can be achieved, 100% crystallization can be realized. We confirmed this with a thermal conductance of $7.81 \times 10^4$ W/m$^2$K (with an average cooling rate of about 12.88 K/ns), getting 20/20 crystallization possibility (see supplementary information Fig. S2).

Furthermore, three representative cases (cases 1, 3, and 10) were chosen to be analyzed by the BOP method to get a better understanding of different temperature evolutions. Fig. 5 (b) shows the central location and the number of atoms of the biggest cluster during crystallization of the three representative cases. At 40 ns, the percentage of crystalline atoms in cases 1, 3, and 10 is about 14%, 55%, and 77%, respectively. The final crystalline structures of the three cases are shown in Fig. 6 (a). There is no well-defined crystalline structure in case 1 (without a clear temperature jump), indicating the atoms are in an amorphous structure (a-Si). In case 3, several crystalline clusters formed with a slower temperature jump compared to the steep temperature jump in case 10, indicating the formation of polycrystalline (pc-Si). Case 10 formed a single-crystal-like structure (sc-Si) with a steep temperature jump. In Fig. 5 (b), the biggest cluster appeared in a spherical shell with a thickness of 5 Å below the surface of the nanoparticle, which is consistent with the previous report.[21] Before the temperature jumps, the number of atoms in the biggest cluster is small. With the temperature jump starting at 10.48 ns, the position of the biggest cluster moved from the sub-surface area (5 Å below the surface) toward the center of the nanoparticle, and the number of crystalline atoms increased rapidly. The cluster position of the pc-Si case and sc-Si case moved close to the center of the nanoparticle, but the a-Si case did not. It is observed that the crystal growth rate associated with the formation of a single-crystal-like structure (case 10, 1575.5 atoms/ns) is substantially higher than that leading to a polycrystalline structure (case 3, 142.5 atoms/ns), while the case resulting in an amorphous structure (case 1) exhibits the lowest growth rate (15 atoms $ns^{-1}$) during the 10–12 ns interval (see supplementary information Fig. S4). In addition, the increase in the number of crystalline atoms correlates with the temperature jump. Therefore, the crystallization process begins at the sub-surface and gradually extends inward to the center of the nanoparticle with latent heat release (more discussion about nucleation location is presented in Section 3.3). We further found that a higher $\Delta T$ indicates a large crystalline atom percentage.

To further gain insight into single-crystal formation, we monitored the atomic structure evolution at different cooling stages over time. As shown in Fig. 6 (b), a small 6-ring structure appears at 8.32 ns near the surface and then dissolves. Then a bigger crystalline cluster shows up again, but in a different location at 9.76 ns. Similar fluctuation repeats until a critical nucleus forms at 10.48 ns, corresponding to the rise in temperature. The nucleation and dissolving near the surface are consistent with the fluctuation of the cluster center position shown in Fig. 5 (b), associated with small crystalline atom numbers (typically in the order of tens). However, upon the formation of nuclei that exceeds the critical size, it becomes thermodynamically stable, and fast crystal growth is observed. Interestingly, a second crystalline cluster is observed at 11.76 ns. This secondary nucleation event occurs after the growth of the first nucleus, during which latent heat release locally increases the temperature and reduces the degree of supercooling. As a result, subsequent nucleation is favored in regions that remain sufficiently undercooled. Afterwards, this second crystal connected and merged with the initial crystal. When the temperature reaches the highest value at 12.56 ns, about 55% of the atoms are crystallized into a single crystalline cluster. During this process, we can observe the cluster center moves from the sub-surface to the center of the particle (also observed in Fig. 5 (b)). Afterward, the crystal continues growing at a slow speed until we stopped at 40 ns. Surprisingly, the nanoparticle shows a single-crystal-like structure (Fig. 6 (b)).

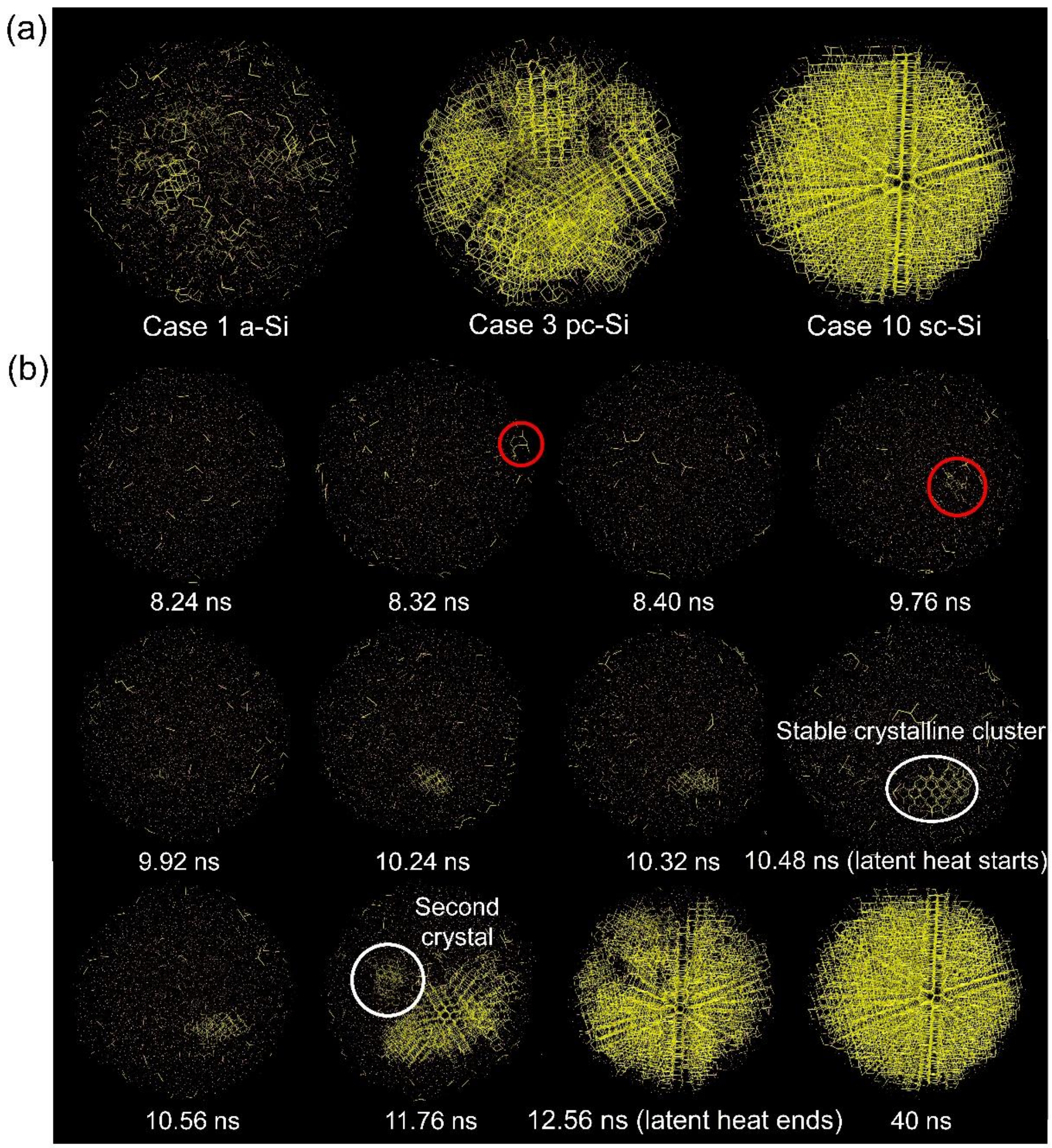

Fig. 6 Crystallization snapshots of an 8 nm Si nanoparticle. (a) Representative atomic structures of amorphous, polycrystalline, and single-crystalline particles at 40 ns. (b) Nucleation and crystal growth probing in the single-crystalline particle. (Red circle indicates the unstable ring structure formed on the surface. White circle indicates stable nucleus.)

To further verify the single-crystal-like structure and monitor the structure evolution, the radial distribution function (RDF) and coordination numbers are calculated for the 8 nm particle. Fig. 7 (a) shows the RDF of amorphous (case 1), polycrystalline (case 3), and single crystal (case 10) structures at 40 ns and compares them to the perfect crystal. According to the broad and low first peak and the absence of long-range peaks, case 1 remains amorphous and does not have a well-crystallized atomic structure. Cases 3 and 10 have sharper peaks, and they have good matches with the perfect diamond structure. Compared to the relatively low peaks of case 3, the sharp peaks of case 10 suggest a better crystalline structure and indicate the possibility of single-crystal formation, which is consistent with the previous report.[26] We further probed the structure evolution with RDF for case 10. As shown in Fig. 7 (b), at the early stage, the first peak is broad, and there is no clear long-range peak. Gradually, peaks start forming from 10.48 ns to 12.56 ns, especially in the long-range area, demonstrating the atomic clustering. When we stopped the simulation at 40 ns, sharp peaks emerged, confirming long-range crystalline order. From 0 to 40 ns, we observe peak formation and then transition from broad to sharp peaks (i.e., a clear increase of peak height), which match well with our above BOP analysis shown in Figs. 5 and 6. Besides, the average coordination number was also used to probe the structure evolution for the 8 nm case 10. To calculate the coordination number, the Si nanoparticle was sliced into 20 concentric shells with each of 2 Å. As shown in Fig. 7 (c), from 1694 K (at 0 ns) to 801 K (at 10.48 ns), the coordination number decreased gradually, approaching the value of a perfect diamond.[37] During the latent heat release, a clear drop in coordination number is observed, which is consistent with the RDF analyses discussed previously. When the simulation ended at 40 ns, the coordination number matched well with the perfect diamond, except for the shells beyond 32 Å. Through checking the visualization of crystallization analyzed by the BOP method in Fig. 6, we found that atoms located in the outer layer of the nanoparticle are difficult to be identified as crystalline atoms. This surface layer (~5 to 8 Å) is well-known for its unique property different from inner atoms, such as the density and potential energy.[21,37] Based on the decrease in coordination number at the surface (Fig. 7c), we can infer that the surface atoms lack sufficient neighboring atoms for accurate analysis, which leads to the mismatch of the surface layer with the perfect diamond. However, an interesting phenomenon was found that the coordination number near the 32 Å shell slightly increases compared with the inner shell (from 16-30 Å), as indicated in Fig. 7c, when the temperature cools down to 1024 K at 10.48 ns. This phenomenon became more pronounced when the temperature jumped to 1373 K (at 12.46 ns) during cooling due to latent heat. When the temperature cools down to 801 K, the coordination number of all the sphere shells is close to a perfect diamond. Combined with the previous BOP analysis, it was found that the region where the coordination number increases is exactly the area where crystallization occurs first.

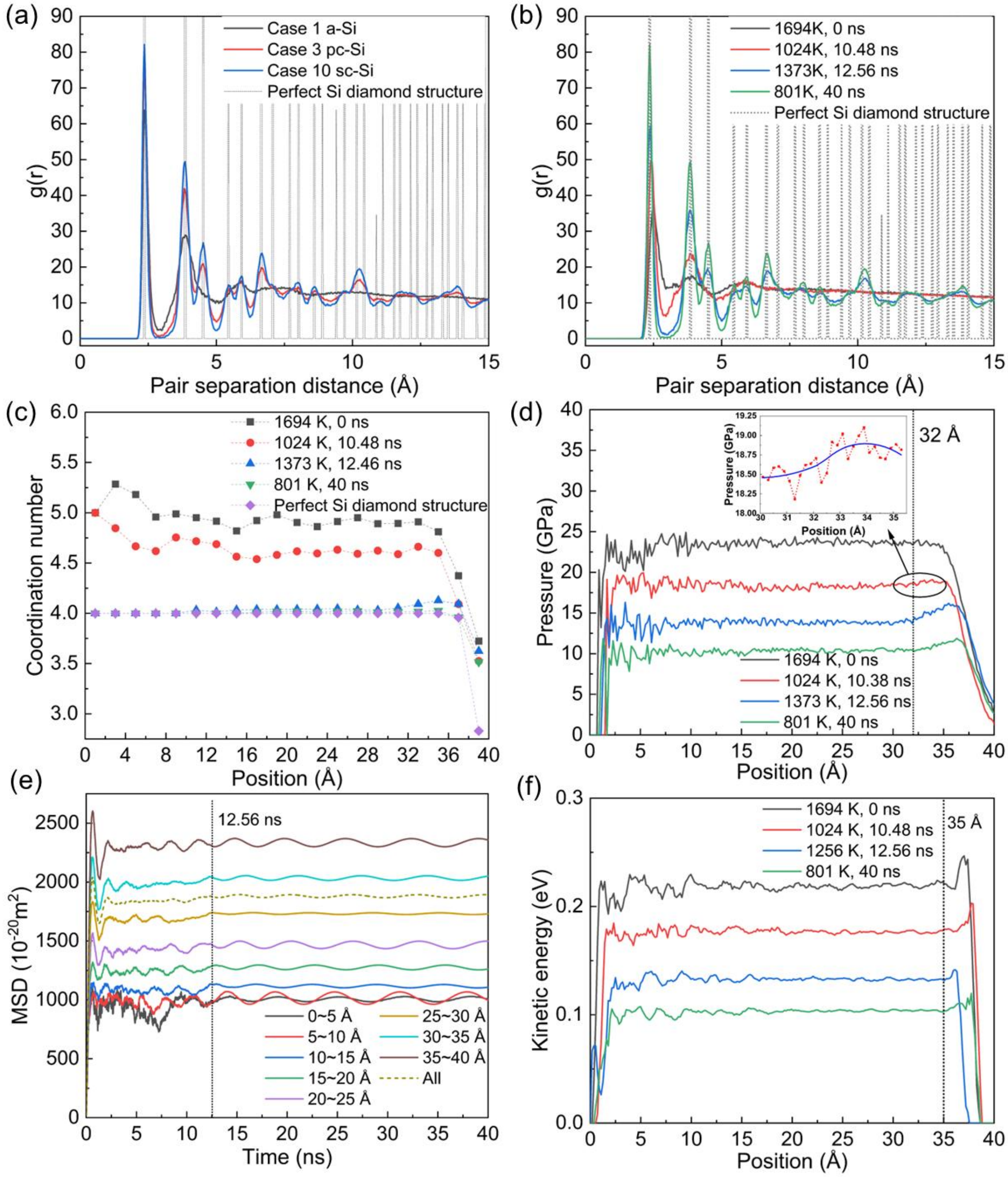

Fig. 7 Structure evolution of an 8 nm Si nanoparticle during solidification. (a) Radial distribution function (RDF) of the representative cases 1, 3, and 10, in comparison with perfect Si diamond structure. (b) RDF of case 10 at different stages. (c) Coordination number of case 10 at different stages. (d) Pressure of case 10 at different stages. (e) Mean squared displacement (MSD) of different shells. (f) Kinetic energy of different stages.

In addition, the pressure distribution from the inner to outer layer of the 8 nm nanoparticle was also calculated (see supplementary information Fig. S6) to examine if the pressure has a direct impact on the initial stable nucleation formation.[19] When the latent heat starts, a slight increase in pressure is observed near the 32 Å region compared with the adjacent inner region (Fig. 7d). The inset in Fig. 7d provides a clearer view of the pressure increase. This pressure difference becomes more obvious after latent heat release (~1073 K, at 12.56 ns) until the final stage at 40 ns. Based on these observations, as well as the density distribution shown in the supplementary information (Fig. S5), we confirm the surface layering phenomenon as reported in previous research.[21,38] We suspect these evidences are closely associated with the observation that the earliest stable nucleation occurs at the sub-surface area in the nanoparticle, similar to previous reports.[19,21,38]

Furthermore, we calculated the mean squared displacement (MSD) and show the results in Fig. 7 (e). It is evident that the atoms in the surface layer (0-5 Å) are more active and move faster (with the highest MSD values) throughout the cooling. We also notice that, after latent heat release (at 12.56 ns, indicated by the vertical dashed line), all the MSD curves become smoother, corresponding to the phase transition from liquid to solid. Here, the fast-moving speed correlates to high kinetic energy as shown in Fig. 7 (f), which indicates the surface atoms might be hotter, or still liquid-like.[20] In addition, the enhanced mobility of surface atoms can also be attributed to weaker atomic bonding and reduced coordination at the surface,[39] which lowers the energy barriers for atomic motion.

### 3.3 Stable nucleation location inside the particle

As discussed in the previous section, crystal growth tends to happen inside the surface layer; thus, in this section, we further monitored and analyzed the formation of the critical nucleus and its transportation. We found that the stable crystalline cluster (~200 atoms) almost always starts from the sub-surface (at least 5 Å away from the surface), regardless of nanoparticle diameter (as shown in Fig. 8a). In contrast, small, unstable crystalline atoms can form anywhere. As suggested by MSD, coordination number analysis, and previous reports, a surface layer,[21] different from bulk Si, exists in nanoparticles, which supports our observation. The critical cluster size is around 200 atoms based on our simulations (at around 0.64 $T_m$), which is close to the reported value of 175 at 0.75 $T_m$ and in line with previous results of 2 nm diameter as the critical nucleus size.[34,40,41] As the size of the critical nucleus is about 200 atoms, it is similar to the total number of atoms of a 2 nm nanoparticle. It may be another reason why 2 nm cannot form a stable crystalline structure. The atomic pressure of the 2 nm nanoparticle is significantly higher than that of the 8 nm nanoparticle (see supplementary information, Fig. S6). Besides, according to previous research, the Laplace pressure (or surface tension) and internal pressure of the 2 nm droplet are significantly higher than those of bigger-sized nanoparticles,[20,42] which indicates that excessive pressure may suppress the crystallization in 2 nm nanoparticles.

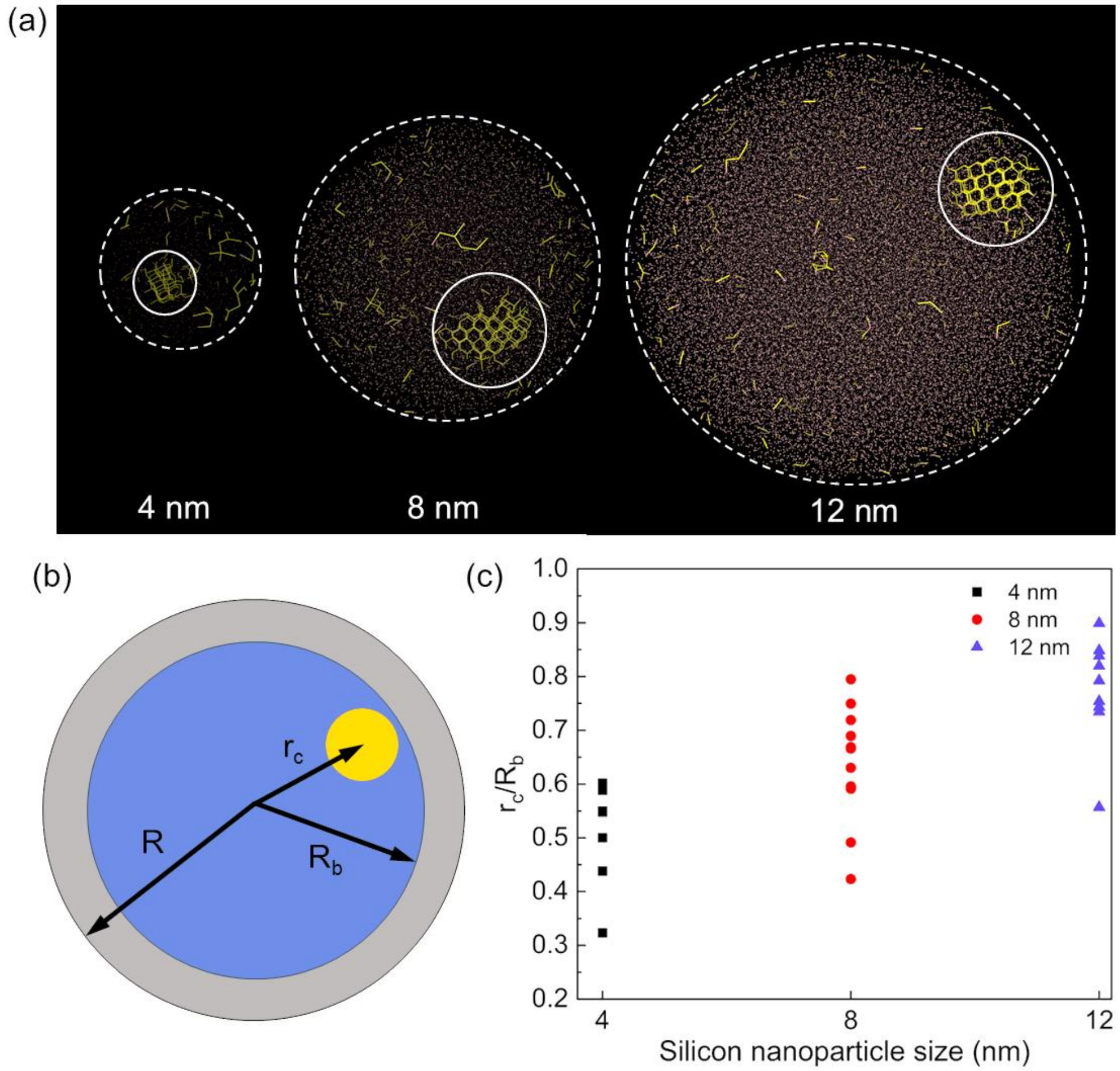

Fig. 8 Stable nucleus location. (a) Critical nucleus location in nanoparticles with different diameters. (b) Schematic drawing of the critical nucleus inside the particle. R is the nanoparticle radius, $R_b$ is the radius of the bulk atoms excluding surface layer atoms, $r_c$ is the distance from the nucleus center to the particle center. (c) Distribution of $r_c/R_b$ for 4, 8, and 12 nm nanoparticles.

For the cases exhibiting crystallization, the distribution of stable cluster centers is plotted in Fig. 8 (c). As the nanoparticle size increases, the probability of larger values of $r_c/R_b$ increases, indicating that sub-surface nucleation (beneath 5 Å) becomes increasingly favored in larger nanoparticles. It is generally believed that during cooling, the surface tends to have a more severe density fluctuation,[21,38] which can lead to a negative pressure to reduce the interfacial energy barrier for nucleation.[19] On the other hand, with the increase of nanoparticle diameter, the corresponding Laplace pressure and surface tension decrease,[20,38] making the surface nucleation relatively more favorable. This result is generally consistent with surface-induced crystallization of Si.[19,21].

## 4. Conclusion

Molecular dynamics simulations were conducted to understand the solidification process of laser-induced forward transfer printing of Si. The droplet size and cooling effects are systematically investigated, which are critical for tuning the resulting atomic structures. The structure evolution during cooling is probed with several techniques (including BOP, coordination number, and MSD), indicating (1) the surface layer is not suitable for stable nucleation formation in nanoparticles; (2) the sub-surface is the position where stable crystallization first occurs; and (3) single-crystal formation is possible. Based on our simulation study, we believe the cooling rate is the most critical parameter to determine the crystallization, which is a function of thermal conductance and nanoparticle size. Generally, we found that with the decrease of cooling rate, nanoparticles tend to have a higher chance to form crystalline structures. Particles with a size of larger than 2 nm can form crystalline and even single-crystal structures with a cooling rate on the order of 10 K/ns. In experiment, compared with femtosecond and picosecond laser processing, which typically has a cooling rate on the order of $10^3$-$10^5$ K/ns, and 10-$10^3$ K/ns,[43–45] nanosecond laser processing offers a comparable cooling rate on the order of 0.01-10 K/ns,[46–49] which is close to our simulated cooling rate. Therefore, we expect, nanosecond laser can be a promising choice for direct printing of crystalline Si structures. We hope this information can guide better experimental design for additive manufacturing of Si and its devices.

## 5. ACKNOWLEDGMENTS

The authors would like to thank the financial support from the National Science Foundation under Grant No. 2531597. The authors would also like to thank the computing resources provided by the Arkansas High Performance Computing Center (AHPCC).